\documentclass[prd,twocolumn,aps,showpacs,nofootinbib,nobibnotes,superscriptaddress,preprintnumbers]{revtex4-1}
\usepackage{epsfig}
\usepackage{graphicx}
\usepackage{bm}
\usepackage[usenames, dvipsnames]{color}
\usepackage{dcolumn}   
\usepackage[spanish,english]{babel}
\usepackage{bm}     
\usepackage{bbm}       
\usepackage{amssymb}  
\usepackage{amsmath}
\usepackage{latexsym}
\usepackage{float}
\usepackage{ifthen}
\usepackage{caption,subfig}
\usepackage{enumerate}
\usepackage{url}
\usepackage{caption,subfig}
\usepackage{amsopn}
\usepackage[colorlinks,citecolor=blue,linkcolor=blue,urlcolor=blue]{hyperref}

\usepackage{amsfonts}
\usepackage{multirow}
\usepackage{array}
\usepackage{booktabs}
\usepackage{rotating}
\usepackage{comment}

\usepackage{ulem}
\def\clap#1{\hbox to 0pt{\hss#1\hss}}

\def\bea{\begin{eqnarray}}
\def\eea{\end{eqnarray}}
\def\be{\begin{equation}}
\def\ee{\end{equation}}

\newcommand{\dd}{{\rm d}}

\newcommand{\QQ}{\mathbb{Q}}

\begin{document}
%\hspace{5.2in} \mbox{NORDITA-2015-38}\\\vspace{1.53cm} % Preprint number

\title{ADM formulation and Hamiltonian analysis of Coincident General Relativity}

\author{Fabio D'Ambrosio} \email{fabioda@phys.ethz.ch}
\author{Mudit Garg} \email{gargm@student.ethz.ch}
\author{Lavinia Heisenberg} \email{lavinia.heisenberg@phys.ethz.ch}
\author{Stefan Zentarra} \email{szentarra@phys.ethz.ch}
\affiliation{Institute for Theoretical Physics,
ETH Zurich, Wolfgang-Pauli-Strasse 27, 8093, Zurich, Switzerland}

\date{\today}

\begin{abstract}
We consider a simpler geometrical formulation of General Relativity based on non-metricity, known as Coincident General Relativity. We study the ADM formulation of the theory and perform a detailed Hamiltonian analysis. We explicitly show the propagation of two physical degrees of freedom, as it should, even though the role of boundary terms and gauge conditions is significantly altered. This might represent an alternative promising new route for numerical relativity and canonical quantum gravity. We also give an outlook on the number of propagating degrees of freedom in non-linear extension of non-metricity scalar.
\end{abstract}

%\pacs{95.35.+d, 04.50.Kd}
%PACS NEEDED

\maketitle

%-----------------------------------------------------------------

\section{Introduction}
%%%%%%%%%%%%%%%%%%%%%%%%%%%%%%%%%%%%%%%%%%%%%%%%%%%%%%%%%%%%%%%%%%
%%%%%%%%%%%%%%%%%%%%%%%%%%%%%%%%%%%%%%%%%%%%%%%%%%%%%%%%%%%%%%%%%%
General Relativity (GR) is the most successful classical theory of gravitation that we have at hand. Based on the equivalence principle and Special Relativity, Einstein postulated that gravitation, caused by any form of energy, manifests itself as the curvature of spacetime. Given a distribution of matter, the Riemannian geometry of spacetime is determined by the Einstein field equations. This theory has withstood intense observational and experimental scrutiny and its most spectacular predictions, such as the existence of black holes and gravitational waves, have been confirmed, opening up a new era in GR and relativistic astrophysics. Within its interpretation as the geometrical property of spacetime, the question arises whether equivalent geometrical formulations of GR exist. In fact, the much richer structure of the affine sector allows two additional and fully equivalent formulations of GR based on flat geometries, attributing gravitation to either torsion~\cite{Aldrovandi:2013wha} or non-metricity~\cite{BeltranJimenez:2017tkd}. This geometrical trinity of gravity~\cite{BeltranJimenez:2019tjy} (see also~\cite{Heisenberg:2018vsk}) offers new perspectives on the computation of gravitational energy-momentum, the entropy of black holes, and on canonical quantization~\cite{Jimenez:2019yyx,BeltranJimenez:2018vdo}.

One of the most important fundamental theoretical developments in GR was its Hamiltonian formulation. In standard mechanics, the Hamiltonian of a system is simply the Legendre transform of the system's Lagrangian function, $H := \sum_{i}p_i \dot q^{i} - L$. This transformation eliminates velocities and renders the Hamiltonian a function of position variables $q^{i}$ and conjugate momenta $p_i$. All equations of motion that follow from a variational principle can be formulated as equivalent Hamiltonian equations, which are really just dynamical equations for $q^{i}$ and $p_i$. Similar considerations are true for field theories and GR can indeed be cast into a Hamiltonian form. However, the transition from a Lagrangian to a Hamiltonian formulation is now more subtle due to the presence of constraints and Dirac's procedure~\cite{Dirac} for generalized Hamiltonian systems has to be employed. Even then, special care is needed due to the theory's general covariance. 

Significant progress in understanding GR's Hamiltonian formulation has been made by the pioneering work of Arnowitt, Deser and Misner (ADM)~\cite{Arnowitt:1959ah} who introduced a new set of variables, which nowadays form the bedrock of numerical relativity. In this formalism, spacetime is foliated into a family of spacelike hypersurfaces $\Sigma_t$ and the metric $g_{\mu\nu}$ can be conveniently decomposed into a lapse function $N$, a shift vector field $N^{i}$, and a three-dimensional spatial metric $\gamma_{ij}$. While lapse and shift enter into the Hamiltonian formulation as Lagrange multipliers, the spatial metric and its conjugate momentum govern the dynamics of GR. These variables allow to cast Einstein's field equations in the form of Hamilton's equations. Apart from numerical relativity, the ADM formalism is at the heart of studies concerning canonical quantum gravity and the causal structure of GR.

The Dirac procedure of constrained systems relies on the degeneracy $\det \mathcal{H}^{\mu\nu}=0$ of the Hessian matrix $\mathcal{H}^{\mu\nu}=\frac{\delta^2\mathcal{S}}{\delta\dot{\mathcal{O}}_\mu\delta\dot{\mathcal{O}}_\nu}$, where $\mathcal{O}$ represents the configuration fields of the theory under consideration. The degeneracy of the Hessian matrix in constrained systems translates into the fact that some of the degrees of freedom do not propagate. In the language of the Hamiltonian formulation, the Hessian prevents the invertibility of configuration space into phase space and its degeneracy restricts the physical phase space to a constrained surface. In order to study how many constraints there are, we need to study the full constraints algebra and the nature of the constraints. So-called first-class constraints point toward the existence of gauge symmetries and remove twice the number of degrees of freedom. The ADM decomposition of standard GR reveals that lapse and shift generate such first-class constraints, thereby removing eight degrees of freedom in configuration space and hence giving rise to two propagating degrees of freedom. 

The study of black holes, gravitational waves, neutron stars, and other similarly strong gravitational phenomena rely strongly on numerical relativity, which heavily uses the ADM $3+1$ decomposition. In this way, Einstein's field equations are treated as a constrained initial value problem, where the initial gravitational fields on some hypersurface are numerically evolved to neighboring hypersurfaces. Stability and convergence of the numerical solutions are indispensable for such a numerical analysis and therefore much effort has been given to coordinates, gauge conditions and reformulations of Einstein's equations in terms of adequate variables. In this Letter we perform a detailed ADM decomposition and Hamiltonian analysis of a representation of General Relativity based on non-metricity, known as Coincident General Relativity (CGR) ~\cite{BeltranJimenez:2017tkd}. Since the role of boundary terms and gauge conditions in CGR are altered, this representation could offer an alternative new route for numerical relativity and canonical quantum gravity.  In addition, we give an outlook on the number of propagating degrees of freedom in $f(\mathcal Q)$.

%%%%%%%%%%%%%%%%%%%%%%%%%%%%%%%%%%%%%%%%%%%%%%%%%%%%%%%%%%%%%%%%%%
%%%%%%%%%%%%%%%%%%%%%%%%%%%%%%%%%%%%%%%%%%%%%%%%%%%%%%%%%%%%%%%%%%
\section{Coincident General Relativity}
In this work we consider an exceptional class of symmetric teleparallel theories of gravity which are consistent with a vanishing affine connection. In this geometrical formulation of GR, gravity is deprived of any inertial character and purged from the boundary terms. The essential starting point is the condition of teleparallelism. Vanishing curvature,
\begin{equation}
R^\alpha{}_{\beta\mu\nu}=2\partial_{[\mu}\Gamma^\alpha{}_{\nu]\beta}+2\Gamma^\alpha{}_{[\mu|\lambda|} \Gamma^\lambda{}_{\nu]\beta}\overset{!}{=}0\;,
\label{eq:zeroR}
\end{equation}
forces the affine connection to have the form
\begin{equation}
\Gamma^\alpha{}_{\mu\beta}=(\Lambda^{-1})^\alpha{}_\rho\partial_\mu\Lambda^\rho{}_\beta,
\label{eq:Telecon}
\end{equation}
where $\Lambda^\alpha{}_\beta\in GL(4,\mathbb{R})$. The second requirement of the symmetric teleparallelism is the vanishing of the torsion tensor,
\begin{equation}
T^\alpha{}_{\mu\beta}=2\Gamma^\alpha{}_{[\mu\beta]}=2(\Lambda^{-1})^\alpha{}_\rho\partial_{[\mu}\Lambda^\rho{}_{\beta]}\overset{!}{=}0.
\label{eq:zeroT}
\end{equation}
This further restricts the form of the connection to be a pure diffeomorphism, 
\begin{equation}
\Gamma^\alpha{}_{\mu\beta}=\frac{\partial x^\alpha}{\partial \xi^\lambda}\partial_\mu\partial_\beta\xi^\lambda \qquad \text{and} \qquad
\Lambda^\alpha{}_\beta=\partial_\beta\xi^\alpha,
\label{GammaSTEGR}
\end{equation}
with arbitrary $\xi^\alpha$. As it becomes clear from the above expression, the gauge choice $\xi^\alpha=x^\alpha$, known as coincident gauge~\cite{BeltranJimenez:2017tkd}, trivializes the connection $\Gamma^\alpha{}_{\mu\beta}=0$. 
The fundamental geometric object of CGR is the non-metricity tensor
\begin{equation}
	Q_{\alpha\mu\nu} := \nabla_\alpha g_{\mu\nu}=\partial_\alpha g_{\mu\nu}-2(\Lambda^{-1})^\lambda{}_\rho\partial_\alpha\Lambda^\rho{}_{(\mu} g_{\nu)\lambda}.
\end{equation}
At quadratic order, there are five independent contractions of the non-metricity tensor,
\begin{align}\label{eqSIIgen}	
\QQ& :=
c_{1}Q_{\alpha\mu\nu}Q^{\alpha\mu\nu} +
	c_{2}Q_{\alpha\mu\nu}Q^{\mu\nu\alpha} +
	c_{3}Q_{\mu}Q^{\mu} \nonumber\\
	&+
	c_{4}\bar{Q}_{\mu}\bar{Q}^{\mu}+
	c_{5}Q_{\mu}\bar{Q}^{\mu},
\end{align}
where the two independent traces are denoted by $Q_\mu:=Q_{\mu\alpha}{}^\alpha$ and $\bar{Q}_\mu:=Q^{\alpha}{}_{\alpha\mu}$. In terms of the non-metricity scalar $\QQ$, the action of symmetric teleparallel gravity can be written compactly as
\begin{equation}
	\mathcal{S}[g, \Gamma] =\int\dd^4 x\, \sqrt{-g}\,\QQ.
\end{equation}
Only for the parameter choices
 \begin{equation}
 c_1=-\frac14, \,c_2=\frac12, \,c_3=\frac14, \,c_4=0, \,c_5=-\frac12,
 \end{equation}
is symmetric teleparallelism equivalent to GR. One can establish a duality relation between the standard Einstein-Hilbert Lagrangian and the quadratic non-metricity Lagrangian,
 \begin{equation}\label{relDual}
 \mathcal{R}={\mathcal Q}-D_\alpha (Q^\alpha-\bar{Q}^\alpha),
  \end{equation}
  where ${\mathcal Q} := \QQ|_{c_i\to\text{GR}}$, the curly Ricci scalar $\mathcal R$ is associated to the Levi-Civita connection of $g$ and $D_\alpha$ is the metric-compatible covariant derivative.
  For later convenience, let us rewrite the CGR action as
\begin{align}\label{actionCGR}
    \mathcal{S}[g, \Gamma] &=\int\dd^4 x\, \sqrt{-g}\,\mathcal Q\notag\\
    &= \frac{1}{4}\int \dd^4 x\,\sqrt{-g}(-g^{\alpha\rho}g^{\beta\mu}g^{\sigma\nu}+2g^{\alpha\nu}g^{\beta\mu}g^{\sigma\rho}\notag\\
    &+g^{\alpha\rho}g^{\beta\sigma}g^{\mu\nu}-2g^{\alpha\beta}g^{\mu\nu}g^{\sigma\rho})Q_{\alpha\beta\sigma}Q_{\rho\mu\nu}.
\end{align}
The theory described by~\eqref{actionCGR} has been shown to be completely equivalent to GR~\cite{BeltranJimenez:2017tkd} and hence represents a new, alternative way of studying the Hamiltonian formulation of GR, with potentially important implications for both numerical relativity and canonical quantum gravity.

%%%%%%%%%%%%%%%%%%%%%%%%%%%%%%%%%%%%%%%%%%%%%%%%%%%%%%%%%%%%%%%%%%
%%%%%%%%%%%%%%%%%%%%%%%%%%%%%%%%%%%%%%%%%%%%%%%%%%%%%%%%%%%%%%%%%%
\section{ADM decomposition}\label{ADM}
In this section we perform the $3+1$ decomposition of the CGR action~\eqref{actionCGR} in terms of ADM variables. We work exclusively in the coincident gauge, $\Gamma^\alpha{}_{\mu\beta}=0$. Consequently, the non-metricity tensor simplifies to $Q_{\alpha\mu\nu}=\partial_\alpha g_{\mu\nu}$ and the action~\eqref{actionCGR} becomes purely a functional of the metric. Furthermore, one can easily verify that the coincident gauge and the ADM decomposition are mutually compatible. We can therefore follow the standard ADM formalism and foliate spacetime by spacelike hypersurfaces $\Sigma_t$ of constant time $t$. Then, we choose the ADM decomposition in which the four dimensional metric $g_{\mu\nu}$ splits in the usual way,
\begin{equation}\label{4.3}
    g_{\mu\nu} = \begin{pmatrix}
                    -N^2+N_iN^i&  && N_i\\
                    N_j & && \gamma_{ij}
                \end{pmatrix},
\end{equation}
where $N$ denotes the lapse function, $N^{i}$ the shift vector field, and $\gamma_{ij}$ the three dimensional spatial metric induced on $\Sigma_t$. While the latter describes the intrinsic geometry of each hypersurface, lapse and shift determine how these hypersurfaces are connected to each other. More precisely, the shift vector field measures how much a given trajectory at constant spatial coordinates is non-orthogonal to the hypersurface and the lapse denotes the proper time per unit coordinate time measured by an observer moving orthogonal to the slices. Since the spatial metric is symmetric, it contains six independent components, the shift vector has three components and the lapse function only one, giving a total of ten independent components for the metric $g_{\mu\nu}$, as expected. Imposing $g^{\mu\alpha}g_{\alpha\nu}=\delta^\mu_{~\nu}$, the components of the inverse metric are easily determined to be $g^{00}=-\frac{1}{N^2}$, $g^{0i}=\frac{N^i}{N^2}$ and $g^{ij}=\gamma^{ij}-\frac{N^iN^j}{N^2}$. The determinant of the four dimensional metric simply decomposes into $\sqrt{-g}=N\sqrt{\gamma}$, with $\gamma$ denoting the three dimensional metric determinant.

We further introduce a vector $n^\mu$ which is orthogonal to the leaves of the foliation and which satisfies the normalization condition $n^\mu n_\mu = -1$. Its components are then found to be $n^\mu = \left(1/N,-N^i/N\right)$ while the components of its associated one-form are given by $n_\mu~=~(-N,\vec{0})$. Since $\gamma_{ij}$ lives on the spatial hypersurfaces we have $\gamma^{\alpha\beta}n_\beta=0$, fulfilling the relation
\begin{equation}\label{4.4}
    \gamma^{\mu\nu} = g^{\mu\nu}+n^\mu n^\nu.
\end{equation}
Using the normal vector, we can introduce the extrinsic curvature of the hypersurface,
\begin{equation}\label{4.5}
    K_{ij} := \frac{1}{2N}(\mathcal D_iN_j+\mathcal D_jN_i-\dot{\gamma}_{ij}),
\end{equation}
where $\mathcal D_i$ is the metric compatible covariant derivative of~$\gamma_{ij}$. 
The extrinsic curvature provides information on how the hypersurface is curved with respect to the manifold in which it is embedded.
With these geometrical quantities, we are well equipped to perform the $3+1$ decomposition of the CGR action and express it in terms of ADM variables,
    \begin{align}\label{4.6}
     &\mathcal{S}[g] = \int \dd^4 x\, \sqrt{\gamma}N\big\{\overset{(3)}{\mathcal Q}-2\gamma^{in}\gamma^{mj}n^\alpha Q_{\alpha nm} n^\nu Q_{ji\nu}\nonumber\\
        &+\frac{1}{4}\tilde{\gamma}^{ijmn}[ n^\alpha Q_{\alpha jm} n^\rho Q_{\rho in} +2n^\beta Q_{i\beta m}n^\mu Q_{j\mu n}\nonumber \\
        &-2Q_{inm}n^\mu n^\nu Q_{j\mu\nu}\nonumber+2n^\alpha Q_{\alpha nm} n^\nu Q_{ji\nu}]\\
        &+2\gamma^{ij}\gamma^{mn}Q_{nij}n^\alpha n^\beta Q_{\alpha\beta m}\nonumber+2\gamma^{ij}n^\alpha n^\beta Q_{\alpha\beta i}n^\mu n^\nu Q_{j\mu\nu} \nonumber\\
        &-2\gamma^{ij}n^\sigma Q_{ij \sigma}n^\rho n^\mu n^\nu Q_{\rho\mu\nu} 
        \big\},
    \end{align}
where $\tilde{\gamma}^{ijmn}=\gamma^{ij}\gamma^{mn}-\gamma^{in}\gamma^{mj}$ and we denoted the three dimensional non-metricity scalar by
\begin{align}
\overset{(3)}{\mathcal Q} :=& \frac{1}{4}(-\gamma^{il}\gamma^{jm}\gamma^{kn}+2\gamma^{in}\gamma^{jm}\gamma^{kl}\nonumber\\
&+\gamma^{il}\gamma^{jk}\gamma^{mn}-2\gamma^{ij}\gamma^{mn}\gamma^{kl})Q_{ijk}Q_{lmn}.
  \end{align}
  As apparent from the above expression of the action~\eqref{4.6}, it is convenient to calculate the projections of the non-metricity tensor along the normal vector. This will allow us to straightforwardly express the action in terms of lapse, shift, the extrinsic curvature, and the intrinsic non-metricity scalar of the hypersurfaces. These projections are explicitly given by
      \begin{align}\label{eq:Projections}
        &n^\alpha  Q_{\alpha jk}=\frac{\gamma_{ik}\partial_jN^i+\gamma_{ij}\partial_kN^i}{N}-2K_{jk},\nonumber\\
        &n^\alpha  Q_{i\alpha k}=\frac{1}{N}\gamma_{jk}\partial_iN^j\nonumber,\\
        &n^\alpha n^\beta  Q_{\alpha\beta k}=\frac{\gamma_{jk}}{N^2}(\dot{N}^j-N^i\partial_iN^j),\nonumber\\
        &n^\alpha n^\beta  Q_{i\alpha\beta }=-\frac{2\partial_iN}{N}\nonumber,\\
        &n^\alpha n^\beta n^\sigma  Q_{\alpha\beta\sigma}=\frac{2}{N^2}(N^i\partial_iN-\dot{N}).
    \end{align}
    Using~\eqref{eq:Projections}, the CGR action~\eqref{4.6} simplifies to
      \begin{align}\label{SCGR}
         {\mathcal S}[g] &= \int \dd^4 x\, \sqrt{\gamma}\Big\{N(\overset{(3)}{\mathcal Q}+K^{ij}K_{ij}-K^2)+K\partial_iN^i\nonumber\\
         &+\dot{N}\left(\frac{\partial_iN^i}{N^2}\right)+\left[\frac{1}{2N^2}(N\gamma^{ij}Q_{kij}-2\partial_kN)\right]\dot{N}^k\nonumber\\
         &+\gamma^{ij}\gamma^{kl}(Q_{jkl}-Q_{kjl})\partial_iN-\frac{\partial_iN^iN^j\partial_jN}{N^2} \nonumber\\
        &+\frac{N^i\partial_jN\partial_iN^j}{N^2}-\frac{\partial_iN^j}{2N}(2\partial_jN^i+N^i\gamma^{mn}Q_{jmn})\Big\}.
    \end{align}
    At this point it is worth mentioning that integrations by parts with subsequent omissions of boundary terms are allowed, without running the risk of altering the symplectic structure of the underlying theory~\cite{Corichi:2014}. We make use of this fact to rewrite the first term in the third line of~\eqref{SCGR}. A simple integration by parts yields
\begin{align}
   & \gamma^{ij}\gamma^{kl}(Q_{jkl}-Q_{kjl})\partial_i N \equiv (\overset{(3)}{Q^i}-\overset{(3)}{\tilde{Q}^i})\mathcal D_iN\nonumber\\
    &=\mathcal D_i[N(\overset{(3)}{Q^i}-\overset{(3)}{\tilde{Q}^i})]-N\mathcal D_i(\overset{(3)}{Q^i}-\overset{(3)}{\tilde{Q}^i})\label{4.8},
\end{align}
where in the first line we introduced $\overset{(3)}{Q^k}:=\gamma^{ij}Q_{kij}$ and $\overset{(3)}{\tilde{Q}^k}:=\gamma^{ij}Q_{ikj}$. We shall subsequently drop the total derivative term on the second line of~\eqref{4.8}. Similarly, the last line of~\eqref{SCGR} can be expressed compactly as 
    \begin{align}
    &\sqrt{\gamma}\left(\frac{N^i\partial_jN\partial_iN^j}{N^2}-\frac{\partial_iN^j}{2N}\left(2\partial_jN^i+N^i\gamma^{mn}Q_{jmn}\right)\right)\nonumber\\
    &=-\partial_iN^j\partial_j\left(\frac{N^i\sqrt{\gamma}}{N}\right).
    \end{align}
    Next, we pay special attention to the second line of~\eqref{SCGR}, where time derivatives acting on lapse and shift appear. The term multiplying the time derivate of the shift vector can be brought into the form $\partial_k(\sqrt{\gamma}/N)$. Analogously, the terms proportional to $K$, $\dot{N}$ and $\partial_jN$ can be brought into the suggestive form
    \begin{align}
    &\sqrt{\gamma}\left(K-\frac{N^j\partial_jN}{N^2}+\frac{\dot{N}}{N^2} \right)=-\partial_\mu(\sqrt{\gamma}n^\mu).
\end{align}
We use all these prearrangements to express our CGR action~\eqref{SCGR} as
\begin{align}
    &{\mathcal S}[g] = \int \dd^4 x\, \bigg\{\sqrt{\gamma}N[\overset{(3)}{\mathcal Q}-\mathcal D_i(\overset{(3)}{Q^i}-\overset{(3)}{\tilde{Q}^i})+K^{ij}K_{ij}-K^2]\label{4.10}\nonumber\\
    &\phantom{ =\int\dd}-\partial_iN^i\partial_\mu(\sqrt{\gamma}n^\mu)+\partial_\mu N^i\partial_i(\sqrt{\gamma}n^\mu)\bigg\}.
\end{align}
In the above expression, the last two terms in the second line are in fact identical, which immediately becomes apparent after performing two partial integrations to swap the derivatives $\partial_\mu$ and $\partial_i$, and dropping the resulting boundary terms. Thus, they cancel each other and the action can be further simplified to
\begin{align}\label{SCGR2}
    \mathcal{S}[g] &=\int \dd^4 x\,\sqrt{\gamma}N[\overset{(3)}{\mathcal Q}-\mathcal D_i(\overset{(3)}{Q^i}-\overset{(3)}{\tilde{Q}^i})+K^{ij}K_{ij}-K^2].
\end{align}
It is worth to emphasize that lapse and shift appear now non-dynamically in this final action, i.e. there are no terms proportional to $\dot N$ and $\dot N^{i}$, respectively. The only dynamical variables are $\gamma_{ij}$, with a priori six degrees of freedom. However, not all of these modes propagate as our Hamiltonian analysis will show in the next section following the standard Dirac procedure.

%%%%%%%%%%%%%%%%%%%%%%%%%%%%%%%%%%%%%%%%%%%%%%%%%%%%%%%%%%%%%%%%%%
%%%%%%%%%%%%%%%%%%%%%%%%%%%%%%%%%%%%%%%%%%%%%%%%%%%%%%%%%%%%%%%%%%
\section{Hamiltonian analysis}\label{sec_Hamiltonian}
In this section we perform the Hamiltonian analysis of the action~\eqref{SCGR2}, following~\cite{Dona:2010}. Let us reiterate that we can always perform partial integrations and remove boundary terms from an action without alteration to the symplectic structure of the theory it describes~\cite{Corichi:2014}. This means that the actions~\eqref{SCGR2} and~\eqref{SCGR} describe the same theory. However, notice that~\eqref{SCGR}, while completely expressed in terms of ADM variables, contains terms proportional to $\dot N$ and $\dot N^{i}$ and would therefore lead to rather complicated primary constraints which contain partial derivatives of field variables. This represents a strong departure from the standard assumptions at the base of Dirac's procedure and complicates the Hamiltonian analysis significantly. 

However, in the action~\eqref{SCGR2} lapse and shift enter non-dynamically and these problems are therefore avoided. In fact, the momentum densities\footnote{We use Ashtekar's tilde notation to indicate tensor densities of weight one, with the exception of Lagrangian and Hamiltonian densities for which we use $\mathcal L$ and $\mathcal H$, respectively.} conjugate to lapse and shift are simply given by $\tilde\pi_N:=\frac{\delta \mathcal{S}}{\delta \dot{N}} =0$ and $\tilde\pi_i:=\frac{\delta \mathcal{S}}{\delta \dot{N}^i}=0$, respectively, and $\tilde\pi_{\mu}:=(\tilde\pi_N, \tilde\pi_i)$ constitute the four primary constraints which define the primary constraint surface $\Gamma_P$.

The only non-vanishing conjugate momentum densities are those associated to the spatial metric,
\begin{align}\label{4.11c}
    \tilde\pi^{ij} :=\frac{\delta \mathcal{S}}{\delta \dot{\gamma}^{ij}}=\sqrt{\gamma}\left(K\gamma^{ij}-K^{ij}\right).
\end{align}
They do not represent constraints since $K_{ij}$ explicitly contains the velocities $\dot{\gamma}_{ij}$. The non-vanishing Poisson brackets\footnote{The equal time PB between two functions $F(x)$ and $G(x)$ of the phase space variables $\{N,N^i,\gamma_{ij},\tilde\pi_N,\tilde\pi_i,\tilde\pi^{ij}\}$ is defined as
\begin{align}
    \{F({x}),G({y})\}&:=\int_{\Sigma_t} \dd^3 z\sum_{k}\left(\frac{\delta F({x})}{\delta \Phi^k({z})}\frac{\delta G({y})}{\delta \tilde\Pi_k({z})}-\frac{\delta G({y})}{\delta \Phi^k({z})}\frac{\delta F({x})}{\delta \tilde\Pi_k({z})}\right),\notag
\end{align}
with the placeholders $\Phi^k=\{N,N^i,\gamma_{ij}\}$ and $\tilde\Pi_k=\{\tilde\pi_N,\tilde\pi_i, \tilde\pi^{ij}\}$.} (PBs) between the phase space variables are
\begin{align}
    \{N({x}), \tilde\pi_N({y})\} &= \delta^{(3)}(\vec{x}-\vec{y}),\notag\\
    \{N^i({x}), \tilde\pi_{j}({y})\} &= \delta^i_{j}\delta^{(3)}(\vec{x}-\vec{y}),\nonumber\\
    \{\gamma_{ij}({x}), \tilde\pi^{mn}({y})\} &= \delta^m_{(i}
    \delta_{j)}^n\delta^{(3)}(\vec{x}-\vec{y}).
\end{align}
This immediately implies that PBs between the primary constraints are strongly zero: $\{\tilde\pi_\mu, \tilde\pi_\nu\}=0$. The primary Hamiltonian density of the system is given by
\begin{align}
    \mathcal{H} := \lambda\,\tilde\pi_N + \lambda^{i}\tilde\pi_i + \tilde\pi^{ij}\dot{\gamma}_{ij}-\mathcal{L},
\end{align}
where $\lambda$ and $\lambda^{i}$ are arbitrary Lagrange multipliers.  
Using $\dot{\gamma}_{ij}=\mathcal D_iN_j+\mathcal D_jN_i-2NK_{ij}$ from the definition of the extrinsic curvature~\eqref{4.5} together with 
\begin{equation}
	K^{ij} = \frac{1}{\sqrt{\gamma}}\left(\tilde\pi^{ij}-\frac12\left(\tilde\pi^{k}_{\ k}\right)^2 \gamma^{ij}\right), 
\end{equation}
which can easily be inferred from~\eqref{4.11c} by taking its trace, the Hamiltonian density can be brought into the form
\begin{align}\label{Ham1}
    \mathcal H =&- \sqrt{\gamma}\Bigg\{N\big[\overset{(3)}{\mathcal Q}-\mathcal D_i(\overset{(3)}{Q^i}-\overset{(3)}{\tilde{Q}^i})\big]\notag\\
    &-\frac{N}{\gamma}\left(\tilde\pi_{ij}\tilde\pi^{ij}-\frac{1}{2}(\tilde\pi^i_{~i})^2\right)+2N^i\mathcal D_j\left(\frac{\tilde\pi^j_{~i}}{\sqrt{\gamma}}\right)\Bigg\}\notag\\
    &+\lambda\,\tilde\pi_N + \lambda^{i}\tilde\pi_i.
\end{align}
Next we introduce the Hamiltonian $H:=\int_{\Sigma_t}\mathcal{H}(x,t)\, \dd^3 x$, which generates evolution in the phase space. In order for this evolution to be consistent with the constraints, the PBs between $\tilde\pi_N$, $\tilde\pi_i$, and $H$ need to vanish on $\Gamma_P$. This leads straightforwardly to the secondary constraints
\begin{align}
    \tilde C_0 &:=- \sqrt{\gamma}\left[\overset{(3)}{\mathcal Q}-\mathcal D_i(\overset{(3)}{Q^i}-\overset{(3)}{\tilde{Q}^i})-\frac{1}{\gamma}\left(\tilde\pi_{ij}\tilde\pi^{ij}-\frac{1}{2}(\tilde\pi^i_{\ i})^2\right)\right] \nonumber\\
    \tilde C_i &:=- 2\mathcal D_j{\tilde\pi^j_{\ i}}.
\end{align}
Notice that these constraints do not depend on lapse and shift and the Hamiltonian can therefore be rewritten in the suggestive form
\begin{align}\label{4.12a}
    H =\int_{\Sigma_t} \dd^3x\, \left(\lambda\,\tilde\pi_N+\lambda^i\tilde\pi_i+N \tilde C_0+N^i \tilde C_i\right),
\end{align}
where lapse and shift are recognized to act as Lagrange multipliers. The simultaneous vanishing of the primary and secondary constraints defines the secondary constraint surface $\Gamma_S\subseteq\Gamma_P$. Demanding that this surface is preserved at all times leads us to check for tertiary constraints. That is, we require the PBs\footnote{Constraints are only imposed after PBs have been computed.}
\begin{align}
     \{\tilde C_0(x), H\} =&\int_{\Sigma_t} \dd^3y\, (N(y)\{\tilde C_0(x), \tilde C_0(y)\}\nonumber\\
     &+N^i(y)\{\tilde C_0(x), \tilde C_i(y)\}),\nonumber\\
    \{\tilde C_i(x), H\} =& \int_{\Sigma_t} \dd^3 y\,(N(y)\{\tilde C_i(x), \tilde C_0(y)\}\nonumber\\
    &+N^j(y)\{\tilde C_i(x), \tilde C_j(y)\})
\end{align}
to vanish on $\Gamma_S$. Notice that the PBs $\{\tilde \pi_\mu, \tilde C_\nu\}$ do not appear in the above expression because they all vanish due to the absence of lapse and shift in the secondary constraints. Hence, our task is to compute the PBs $\{\tilde C_\mu, \tilde C_\nu\}$ in order to check whether there are tertiary constraints. To that end, and in order to avoid delta distributions in our computations, we introduce the smeared scalar and vector constraints\footnote{Note that the smearing is valid for arbitrary test functions, not just for lapse and shift.}
\begin{align}
	C_S(N) &:=\int_{\Sigma_t} N\,\tilde C_0\,\dd^3 x\notag\\
	C_V(\vec{N}) &:=\int_{\Sigma_t} N^i \tilde C_i\, \dd^3 x,
\end{align}
where $\vec{N} := N^i\partial_i$. First of all, the PBs of the vector constraint with the dynamical phase space variables yield
\begin{align}
    \{\gamma_{ij}(x), C_V(\vec{N})\} &= \int_{\Sigma_t}\dd^3 x\,\delta^m_{(i}
    \delta_{j)}^n\frac{\delta C_V}{\delta \tilde\pi^{mn}(x)}\notag\\
    &= \mathcal{L}_{\vec{N}}\gamma_{ij}(x) = 2\mathcal D_{(i} N_{j)} \nonumber\\
    \{\tilde \pi^{ij}(x), C_V(\vec{N})\} &= -\int_{\Sigma_t}\dd^3 x\, \delta_m^{(i}
    \delta^{j)}_n\frac{\delta C_V}{\delta \gamma_{mn}(x)}\notag\\
    &= -2\tilde\pi^{k(i}\mathcal D_k N^{j)}+\mathcal D_k(\tilde\pi^{ij} N ^k)\notag\\
    &= \mathcal{L}_{\vec{N}}\tilde\pi^{ij}(x),
\end{align}
where $\mathcal{L}_{\vec{N}}$ denotes the Lie derivative with respect to the vector field $\vec{N}$. Note that the PBs between the vector constraint ${C_V}(\vec{N})$ and all other phase space variables vanish on the constraint surface $\Gamma_S$. Therefore, the vector constraint generates the spatial diffeomorphisms for any arbitrary function $F$ of $\gamma_{ij}$ and $\tilde\pi^{ij}$,
\begin{align}\label{eq:SDiffGenerator}
    \{F(\gamma_{ij}, \tilde\pi^{ij}),{C_V}(\vec{N})\} = \mathcal{L}_{\vec{N}}F(\gamma_{ij}, \tilde\pi^{ij}),
\end{align}
and treat all other phase space variables as constants. Next, we pay special attention to the PBs between the smeared scalar and vector constraint, making use of~\eqref{eq:SDiffGenerator}:
\begin{align}
	\{C_S(N), C_V(\vec{N})\} &= \int_{\Sigma_t} \dd^3x\, N(x)\{\tilde C_0(x),{C_V}(\vec{N})\}\nonumber\\
    &=-\int_{\Sigma_t} \dd^3 x\, {\tilde C_0}(x)\mathcal{L}_{\vec{N}} N(x)\notag\\
    &=-{C_S}(\mathcal{L}_{\vec{N}} N)\label{PB1},\\
	\{C_V(\vec{N_1}), C_V(\vec{N_2})\} &= \int_{\Sigma_t} \dd^3 x\,N_1^i(x)\{\tilde C_i(x), C_V(\vec{N}_2)\} \notag\\
    &=-\int_{\Sigma_t} \dd^3x\,{\tilde C_i}(x) \mathcal{L}_{\vec{N}_2} N^i_1(x)\notag\\
    &=-C_V(\mathcal{L}_{\vec{N}_2}\vec{N}_1) = -C_V([\vec{N}_2,\vec{N}_1])\notag\\
    &= C_V([\vec{N}_1,\vec{N}_2])\label{PB2},
\end{align}
where we have used integration by parts and the Lie bracket $[\vec{N}_1,\vec{N}_2]$ defined by
\begin{equation}
    [\vec{N}_1,\vec{N}_2]=(N_1^j\partial_j N_2^i- N_2^j\partial_j N_1^i)\partial_i.
\end{equation}
Since $C_S$ and $C_V$ both vanish on $\Gamma_S$ independently of their argument, we find that the above PBs are preserved by the evolution generated by the Hamiltonian $H$.
What remains to be checked is the PB between two smeared scalar constraints. To that end, it is convenient to split $C_S(N)$ into algebraic and non-algebraic parts with respect to $\gamma_{ij}$ and $\tilde\pi^{ij}$:
\begin{align}
    C_S({N}) &= -\int_{\Sigma_t} \dd^3 x\, N\sqrt{\gamma}\Bigg[\underbrace{\overset{(3)}{\mathcal Q}-\mathcal D_i(\overset{(3)}{Q^i}-\overset{(3)}{\tilde{Q}^i})}_{\text{non-algebraic}} \nonumber\\
    &\phantom{=\ }-\underbrace{\frac{1}{\gamma}\left(\tilde\pi_{ij}\tilde\pi^{ij}-\frac{1}{2}(\tilde\pi^i_{\ i})^2\right)}_{\text{algebraic}}\Bigg]\nonumber\\
    &=:{C_S}^{\text{alg}}(N)+{C_S}^{\text{non-alg}}(N).
\end{align}
One immediate observation is that the algebraic part gives $\{{C_S}^{\text{alg}}(N_1),{C_S}^{\text{alg}}(N_2)\}=0$ as well as $\{{C_S}^{\text{non-alg}}(N_1),{C_S}^{\text{non-alg}}(N_2)\} = 0$. The non-trivial contribution to the PB between the scalar constraints comes from
\begin{align}
&\{{C_S}^{\text{alg}}(N_1), C_S^{\text{non-alg}}(N_2)\} + \{{C_S}^{\text{non-alg}}(N_1), C_S^{\text{alg}}(N_2)\} \notag\\
&=-2\int_{\Sigma_t} \dd^3 z\,\tilde\pi_{ij}[N_2\mathcal D^i \mathcal D^j N_1- N_1 \mathcal D^i \mathcal D^j N_2]\nonumber\\
    &=\int_{\Sigma_t} \dd^3 z\, [N_1 \mathcal D^i N_2- N_2 \mathcal D^i N_1]\tilde C_i\notag\\
    &=C_V\left(\left(N_1 \partial^i N_2 - N_2 \partial^i N_1\right)\partial_i\right),
\end{align}
where we have used integration by parts (omitting boundary terms) and in the second line we recognized the smeared vector constraint. Summarizing, we conclude that the PBs between the smeared secondary constraints are given by
\begin{align}
	\{C_S(N), C_V(\vec{N})\} &= -C_S(\mathcal L_{\vec{N}}N)\\
	\{C_V(\vec{N}_1), C_V(\vec{N}_2)\} &= C_V([\vec{N}_1, \vec{N}_1])\notag\\
	\{C_S(N_1),C_S(N_2)\} &= C_V\left(\left(N_1 \partial^i N_2 - N_2 \partial^i N_1\right)\partial_i\right),\notag 
\end{align}
which all vanish on the constraint surface $\Gamma_S$. Hence, there are no tertiary constraints to take into consideration. Moreover, we have shown that $\{\tilde \pi_\mu,\tilde \pi_\nu\} = 0$, $\{\tilde \pi_\mu, \tilde C_\nu\} = 0$, and $\{\tilde C_\mu, \tilde C_\nu\} \approx 0$\footnote{Two phase space functions are said to be weakly equal, $F(q,p)\approx G(q,p)$, if and only if $F(q,p) = G(q,p)$ up to an arbitrary linear combination of constraint.}, which implies that we have in total four first-class primary constraints and four first-class secondary constraints. This removes a total of $2\times 4 + 2\times 4 = 16$ degrees of freedom from the original $20$ degrees of freedom of the unconstrained phase space, leaving only four physical degrees of freedom. Or two physical degrees of freedom in configuration space. As expected, we have the same number of propagating degrees of freedom as in GR.

This is no surprise since the non-metricity scalar $\mathcal Q$ which defines the CGR action satisfies the duality relation~\eqref{relDual} and is therefore equivalent to the Einstein-Hilbert action up to a boundary term. Moreover, one can prove a further duality relation between the intrinsic non-metricity scalar and the intrinsic curvature,
\begin{equation}
	\overset{(3)}{\mathcal Q}-\mathcal D_i(\overset{(3)}{Q^i}-\overset{(3)}{\tilde{Q}^i}) = \overset{(3)}{\mathcal R},
\end{equation}
by virtue of which one recognizes~\eqref{SCGR2} as the ADM action of GR. Hence, CGR possesses the same symplectic structure and propagates the same number of degrees of freedom as GR, but the role of boundary terms and gauge conditions are altered. The physical Hamiltonian of CGR therefore differs from the one of GR and CGR may therefore offer an alternative new route for numerical relativity and canonical quantum gravity~\cite{Jimenez:2019yyx}.\\
\section{Outlook on $f(\mathcal Q)$}
So far we focussed our attention on CGR described by the Lagrangian $\mathcal L = \sqrt{-g}\,\mathcal Q$. However, non-linear extensions of the form $\mathcal L = \sqrt{-g}\, f(\mathcal Q)$ with $f'\neq 0$ have been studied in the literature and interesting consequences for cosmology have been described~\cite{Jimenez:2019ovq, dambrosio:2020}. It is therefore natural to wonder, how many degrees of freedom $f(\mathcal Q)$ theory propagates and how this number depends on the form of~$f$.

At the beginning of the previous section, we emphasized the importance of having removed terms proportional to $\dot N$ and $\dot N^{i}$ from the action in order to proceed with the Hamiltonian analysis. Also in $f(\mathcal Q)$ lapse and shift are non-dynamical. We can venture an educated guess on the propagating number of degrees of freedom based on insights gained in~\cite{Jimenez:2019ovq} from studying cosmological perturbations around a FLRW background in conjunction with the fact that lapse and shift are non-dynamical variables. For later reference, we recall that the metric perturbations assume the form
\begin{align}
	\delta g_{00} &= -2 a^2 \phi\\
	\delta g_{0i} &= \delta g_{i0} = a^2\left(\partial_i B + B_i\right)\notag\\
	\delta g_{ij} &= 2a^2\left[-\psi \delta_{ij} + \left(\partial_i\partial_j - \frac{\delta_{ij}}{3}\partial^k\partial_k\right)E + \partial_{(i}E_{j)}+h_{ij}\right]\notag
\end{align}
and using this ansatz it was shown in~\cite{Jimenez:2019ovq} that $f(\mathcal Q)$ propagates at least two additional degrees of freedom. Since lapse and shift are non-dynamical, they lead to four primary constraints in the Hamiltonian theory. Hence, $f(\mathcal Q)$ could a priori propagate four, five, or six degrees of freedom. However, the perturbation analysis has uncovered that only $\psi$, $E$, $E^{i}$, and $h_{ij}$ can propagate. Hence, in order to obtain five propagating degrees of freedom, either $\phi$ or $B$ need to become dynamical. But this is in contradiction with the fact that lapse and shift are non-dynamical, thus leaving us with the possibility of having four or six degrees of freedom.

Attaining six degrees of freedom would only be possible if the vector perturbation $E^i$, subjected to the constraint $\partial_i E^{i} = 0$, would propagate its two degrees of freedom. Evidence from cosmological perturbation theory suggests that either $E^{i}$ do not propagate at all or there is a strong coupling problem for the cosmological background, due to which kinetic term of $E^{i}$ vanishes~\cite{Jimenez:2019ovq}.

We can consider this counting problem also from a different vantage point and reach the same conclusion: lapse and shift give rise to four primary constraints. These constraints can either be first- or second-class. Excluding the trivial case of GR where all constraints are first-class, and also excluding for the time being the case where all constraints are second-class, leaves us with two possibilities. Either the constraint pertaining to the non-dynamical lapse is first-class, while the constraint pertaining to the shift vector is second-class, or vice versa. In the first case one would obtain $10  - 2\times 1 - 1\times 3 = 5$ while the second case yields $10 - 1\times 1 - 2\times 3 = 3$ degrees of freedom. Both outcomes are in contradiction with the findings from cosmological perturbation theory, strongly suggesting that six degrees of freedom is the only viable option, where all the primary constraints are second class.

A detailed Hamiltonian analysis of $f(\mathcal Q)$ is currently underway in order to further investigate the above arguments.

%%%%%%%%%%%%%%%%%%%%%%%%%%%%%%%%%%%%%%%%%%%%%%%%%%%%%%%%%%%%%%%%%%
%%%%%%%%%%%%%%%%%%%%%%%%%%%%%%%%%%%%%%%%%%%%%%%%%%%%%%%%%%%%%%%%%%
\section{Conclusion}
In this Letter we performed the ADM decomposition and the Hamiltonian analysis of Coincident General Relativity (CGR), thereby providing an independent proof that CGR propagates two physical degrees of freedom. We started with the quadratic non-metricity action in the coincident gauge with vanishing connection and foliated the spacetime into spacelike hypersurfaces. In this $3+1$ decomposition, the metric can be represented in terms of the ADM variables -- lapse, shift, and the spatial metric -- since the coincident gauge and the ADM decomposition are mutually compatible. After making extensive use of projections between the normal vector and the non-metricity tensor, and applying integrations by parts, we accomplished to express the CGR action in terms of lapse, shift, extrinsic curvature, and the non-metricity scalar of the spacelike hypersurfaces. The resulting action does not carry any dynamics for the lapse and shift fields and it can be shown to be equivalent to the ADM action of GR. To complete our analysis and explicitly show that CGR propagates two degrees of freedom, we reviewed the Hamiltonian analysis of GR in ADM variables. We showed that lapse and shift generate first-class constraints which eliminate eight degrees of freedom in configuration space.

This is no surprise since the non-metricity scalar $\mathcal Q$ satisfies the duality relation~\eqref{relDual} with the Ricci scalar. Hence, the CGR action and the Einstein-Hilbert action only differ by a boundary term which has no influence on the symplectic structure of the theory. However, the role of boundary terms and gauge conditions are altered in CGR and the physical Hamiltonian of CGR differs from the one of GR. This difference may offer an alternative new route for numerical relativity and canonical quantum gravity. We leave the exploration of these consequences to future work. 

In the final section of this Letter we also ventured an educated guess about the number of physical degrees of freedom of $f(\mathcal Q)$ theory and found strong indications for six propagating modes.

%\section*{Appendices}

\section*{Acknowledgements}
LH is supported by funding from the European Research Council (ERC) under the European Unions Horizon 2020 research and innovation programme grant agreement No 801781 and by the Swiss National Science Foundation grant 179740. 

%%%%%%%%%%%%%%%%%%%%%%%%%%%%%%%%%%%%%%%%%%%%%%%%%%%%%%%%%%%%%%%%%%
%%%%%%%%%%%%%%%%%%%%%%%%%%%%%%%%%%%%%%%%%%%%%%%%%%%%%%%%%%%%%%%%%%


\begin{thebibliography}{99}

%\cite{Aldrovandi:2013wha}
\bibitem{Aldrovandi:2013wha}
  R.~Aldrovandi and J.~G.~Pereira,
  %``Teleparallel Gravity : An Introduction,''
  Fundam.\ Theor.\ Phys.\  {\bf 173} (2013).

  
%\cite{BeltranJimenez:2017tkd}
\bibitem{BeltranJimenez:2017tkd}
  J.~Beltr\'an~Jim\'enez, L.~Heisenberg and T.~Koivisto,
  %``Coincident General Relativity,''
  Phys.\ Rev.\ D {\bf 98} (2018) no.4,  044048,
  %doi:10.1103/PhysRevD.98.044048
  \href{https://arxiv.org/abs/1710.03116}{arXiv:1710.03116 [gr-qc]}.

  
%\cite{BeltranJimenez:2019tjy}
\bibitem{BeltranJimenez:2019tjy} 
  J.~B.~Jim\'enez, L.~Heisenberg and T.~S.~Koivisto,
  %``The Geometrical Trinity of Gravity,''
  Universe {\bf 5}, no. 7, 173 (2019),
  %doi:10.3390/universe5070173
  \href{https://arxiv.org/abs/1903.06830}{arXiv:1903.06830 [hep-th]}.
 

%\cite{Heisenberg:2018vsk}
\bibitem{Heisenberg:2018vsk} 
  L.~Heisenberg,
  ``A systematic approach to generalisations of General Relativity and their cosmological implications,''
  Phys.\ Rept.\  {\bf 796}, 1 (2019),
  %doi:10.1016/j.physrep.2018.11.006
  \href{https://arxiv.org/abs/1807.01725}{arXiv:1807.01725 [gr-qc]}.

  
%\cite{Jimenez:2019yyx}
\bibitem{Jimenez:2019yyx}
  J.~Beltr\'an~Jim\'enez, L.~Heisenberg and T.~S.~Koivisto,
  %``The canonical frame of purified gravity,''
  Int. J. Mod. Phys. D \textbf{28} (2019) no.14, 1944012,
  %doi:10.1142/S0218271819440127
  \href{https://arxiv.org/abs/1903.12072}{arXiv:1903.12072 [gr-qc]}.


%\cite{BeltranJimenez:2018vdo}
  \bibitem{BeltranJimenez:2018vdo}
  J.~Beltr\'an~Jim\'enez, L.~Heisenberg and T.~S.~Koivisto,
  %``Teleparallel Palatini theories,''
  JCAP \textbf{08} (2018), 039,
  %doi:10.1088/1475-7516/2018/08/039
  \href{https://arxiv.org/abs/1803.10185}{arXiv:1803.10185 [gr-qc]}.


\bibitem{Dirac}
  P. A. M. Dirac,{\it Lectures on Quantum Mechanics}, Yeshiva University Press, New York,1964

%\cite{Arnowitt:1959ah}
\bibitem{Arnowitt:1959ah}
  R.~L.~Arnowitt, S.~Deser and C.~W.~Misner,
  %``The Dynamics of general relativity,''
  Gen. Rel. Grav. \textbf{40} (2008), 1997-2027
  %doi:10.1007/s10714-008-0661-1
  \href{https://arxiv.org/abs/gr-qc/0405109}{arXiv:gr-qc/0405109 [gr-qc]};
  R.~L.~Arnowitt, S.~Deser and C.~W.~Misner,
  %``Dynamical Structure and Definition of Energy in General Relativity,''
  Phys. Rev. \textbf{116} (1959), 1322-1330
  %doi:10.1103/PhysRev.116.1322

  
%\cite{Corichi:2014}
\bibitem{Corichi:2014}
  A.~Corichi, I.~Rubalcava-Garc\'{i}a and T.~Vuka\v{s}inac,
  %``Hamiltonian and Noether charges in first order gravity,''
  Gen. Rel. Grav. \textbf{46} (2014), 1572-9532
  %doi:10.1007/s10714-014-1813-0
  \href{https://arxiv.org/abs/1312.7828}{arXiv:1312.7828 [gr-qc]}


%\cite{Dona:2010}
\bibitem{Dona:2010}
  P.~Don\'{a}, S.~Speziale,
  ``Introductory lectures to loop quantum gravity,'' 
  \href{https://arxiv.org/abs/1007.0402}{arXiv:1007.0402 [gr-qc]}
   

%\cite{Jimenez:2019ovq}
\bibitem{Jimenez:2019ovq} 
  J.~Beltr\'an Jim\'enez, L.~Heisenberg, T.~S.~Koivisto and S.~Pekar,
  %``Cosmology in $f(Q)$ geometry,''
  \href{https://arxiv.org/abs/1906.10027}{arXiv:1906.10027 [gr-qc]}.


%\cite{dambrosio:2020}
\bibitem{dambrosio:2020}
  F.~D'Ambrosio, M.~Garg and L.~Heisenberg,
  %``Non-linear extension of non-metricity scalar for MOND,''
  \href{https://arxiv.org/abs/2004.00888}{arXiv:2004.00888 [gr-qc]}


\end{thebibliography}
\end{document}